# Excess H, Suppressed He, and the Abundances of Elements in Solar Energetic Particles

**Donald V. Reames**

Institute for Physical Science and Technology, University of Maryland, College Park, MD 20742-2431 USA, email: dvreames@umd.edu

**Abstract** Recent studies of the abundances of H and He relative to those of heavier ions in solar energetic particle (SEP) events suggest new features in the underlying physics. Impulsive SEP events, defined by uniquely large enhancements of Fe/O, emerge from magnetic reconnection in solar jets. In small, "pure," shock-free, impulsive SEP events, protons with mass-to-charge ratio $A/Q = 1$ fit the power-law dependence of element abundance enhancements *versus A/Q* extrapolated from the heavier elements $6 \leq Z \leq 56$. Sometimes these events have order-of-magnitude suppressions of He, even though H fits with heavier elements, perhaps because of the slower ionization of He during a rapid rise of plasma from the chromosphere. In larger impulsive SEP events, He fits, but there are large proton excesses relative to the power-law fit of $Z > 2$ ions, probably because associated coronal mass ejections (CMEs) drive shock waves fast enough to reaccelerate the impulsive SEPs but also to sample protons from the ambient solar plasma. In contrast, gradual SEP events are accelerated by wide, fast CME-driven shock waves, but those with smaller, weaker shocks, perhaps quasi-perpendicular, favor impulsive suprathermal residue left by many previous jets, again supplemented with excess protons from ambient coronal plasma. In the larger, more common gradual SEP events, faster, stronger shock waves sample the ambient coronal plasma more deeply, overwhelming any impulsive-ion component, so that proton abundances again fit the same power-law distribution as all other elements. Thus, studies of the power-law behavior in $A/Q$ of SEP element abundances give compelling new information on the varying physics of SEP acceleration and properties of the underlying corona.

Keywords: Solar energetic particles · Solar system abundances · Coronal mass ejections · Solar flares

# 1. Introduction

The *Wind* spacecraft has enabled studies of hundreds of solar energetic-particle (SEP) events in its highly productive 25 years of operation. All of these events are different. One most profound difference lies in the relative abundances of the elements in SEPs compared with corresponding abundances in the solar photosphere or corona. New abundances of H and He add information that is sensitive to the physics of particle selection and acceleration. Some elements in SEP events are preferentially selected for acceleration while others are preferentially scattered as they propagate away from the source.

Two primary sources of SEP events have been identified (*e.g.* Reames, 1988, 1995b, 1999, 2013, 2015, 2017a; Gosling, 1993). In the small "impulsive" SEP events, where acceleration has been traced to sites of magnetic reconnection in solar jets (Kahler, Reames, and Sheeley, 2001; Bučík *et al.*, 2018a, 2018b), element abundances are enhanced as a power-law in mass-to-charge ratio $A/Q$, increasing by a factor of ≈1000 across the periodic table from H and He to Au and Pb (Reames and Ng, 2004; Reames, Cliver, and Kahler, 2014a) probably because of the reconnection physics (*e.g.* Drake *et al.*, 2009).

In contrast, in the large "gradual" SEP events (Lee, Mewaldt, and Giacalone, 2012; Desai and Giacalone, 2016), shock waves, driven by fast, wide coronal mass ejections (CMEs; Kahler *et al.*, 1984; Lee, 1983, 2005; Zank, Rice, and Wu, 2000; Cliver, Kahler, and Reames, 2004; Gopalswamy *et al.*, 2012), mostly sample ambient coronal material. This material differs from that in the photosphere by a factor that depends upon the first ionization potential (FIP) of the element. High-FIP (>10 eV) elements are neutral atoms in the chromosphere while low-FIP elements are ions that are preferentially enhanced by a factor of ≈4, probably by the action of Alfvén waves (Laming, 2015; Reames, 2018a; Laming *et al.*, 2019), when swept up into the corona and later sampled as SEPs (Webber, 1975; Meyer, 1985; Reames, 1995a, 2014). After acceleration, ion scattering, depending upon magnetic rigidity, hence upon $A/Q$ at a given velocity (Parker, 1963; Ng, Reames, and Tylka, 1999, 2001, 2003, 2012; Reames, 2016a, 2019b), can also lead to power-law dependence, *i.e.* since Fe scatters less than O, for example, Fe/O is enhanced early but depleted later in an event, producing power-law enhancements that in-crease and then decrease with $A/Q$ (*e.g.* Breneman and Stone, 1995). The seed population sampled by shock waves often contains remnant ions pre-accelerated in impulsive events (Tylka *et al.*, 2001; Mason, Mazur, and Dwyer, 2002); these impulsive seed particles can even dominate in ≈25% of gradual SEP events (Reames, 2016a).

The dependence upon $A/Q$, and thus upon $Q$, is a dependence upon the source-plasma electron temperature $T$, which allows us to estimate $T$ in a large sample of events, both impulsive (Reames, Meyer, and von Rosenvinge, 1994; Reames, Cliver, and Kahler, 2014b, 2015), where $T \approx 3$ MK, and gradual (Reames, 2016a, 2016b), where $0.8 \leq T \leq, 4$ MK. The technique for making such estimates, by best-fitting enhancements *versus $A/Q$* for many values of $T$ and selecting the one with minimum $\chi^2$, has been described in review articles (Reames, 2018b) and even textbooks (Reames, 2017a).

Thus it is common for SEP events of both types to show a power-law dependence on $A/Q$ of observed element abundances divided by the corresponding reference SEP-coronal abundances listed in the Appendix. It is these reference abundances that differ from solar-photospheric abundances because of the "FIP-effect" (*e.g.* Meyer, 1985; Reames, 1995a, 2014). The coronal abundances deduced from SEPs differ from those found for solar wind (Mewaldt *et al.*, 2002; Desai *et al.*, 2003; Kahler, Tylka, Reames, 2009; Reames, 2018a; Laming, *et al.,* 2019). Current theory (Laming, 2015; Laming, *et al.,* 2019) suggests that the FIP-effect seen by SEPs begins on closed magnetic loops where Alfvén waves can resonate with the loop length, while that of the solar wind occurs on magnetic field lines that are open near the chromosphere-coronal boundary (Reames, 2018a, Laming, *et al.*, 2019). Thus SEPs are *not* accelerated solar wind.

Now we come to the elements He and H. None of the early studies of $A/Q$-dependence of SEP abundances included H, and many omitted He as well. Recent studies have shown that the source abundance of He/O can vary by a factor of ≈2 in gradual SEP events (Reames, 2017b, 2018b) but there are occasionally tenfold suppressions in impulsive SEP events (Reames, 2019a). Theoretically, suppression of He may result from its slow ionization, because of its uniquely high FIP of 24.6 eV, as it traverses the chromosphere (Laming, 2009). But how does this produce such large suppressions?

In contrast, H/O shows a much different behavior. In small impulsive SEP events (Reames, 2019b), H fits well onto the extension to $A/Q$ =1 of the power-law fit of en-hancements *versus A/Q* for the elements with $Z \geq 6$, independently of whether He also fits. For larger impulsive events, and especially those associated with faster, but characteristically narrow, CMEs, the enhancement of H can exceed that fit line by an order of magnitude or more – a significant proton excess. Gradual SEP events also fall into two groups (Reames, 2019c). For those associated with ambient coronal temperatures of 0.8 – 1.5 MK, H generally falls close to the fit line from the elements with $Z \geq 6$ extrapolated down to $A/Q = 1$. This applies to about 60% of gradual events. However, for those gradual SEP events with $T \approx 3$ MK, which all have positive power-law slopes for $Z \geq 6$, presumably derived from shock-reaccelerated impulsive-SEP material, H again exceeds the extended fit line by an order of magnitude or more – a factor-of-ten proton excess.

Why do both impulsive and gradual events have subsets with a strong proton excess? Both of these subsets have element enhancements that increase with *A/Q* and involve shock reacceleration of impulsive-SEP ions. To what extent do these subsets involve different physics? Have we merely made some mistake where we draw the line between impulsive and gradual events? Have we misidentified events? In this article we attempt to display and contrast those properties of the events that help disentangle our understanding of the physics of the relative abundances of H and He in all types of SEP events. What is the overall picture? How do the excesses and suppressions fit in?

Element abundances in this article were measured by the *Low-Energy Matrix Telescope* (LEMT) onboard the *Wind* spacecraft, near Earth (von Rosenvinge *et al.*, 1995; see also Chapt. 7 of Reames, 2017a). LEMT measures elements from H through Pb in the region of 2–20MeV amu$^{-1}$, although energies of H are limited to 2 – 2.5 MeV and LEMT resolves only element groups above Fe as shown by Reames (2000, 2017a). We reexamine the impulsive events listed and studied by Reames, Cliver, and Kahler (2014a, 2014b) and the gradual events listed and studied by Reames (2016a). CME data used are observed by the *Large Angle and Spectrometric Coronagraph* (LASCO) onboard the *Solar and Heliospheric Observatory* (SOHO) as reported in the SOHO/LASCO CME catalog (Gopalswamy *et al.*, 2009; https://cdaw.gsfc.nasa.gov/CME_list/).

## 2. Impulsive SEP Events

The uniqueness of impulsive SEP events was first recognized by the unusual abundance of $^{3}$He. Ratios of $^{3}$He/$^{4}$He as high as $1.5 \pm 0.1$ (*e.g.* Serlemitsos and Balasubrahmanyan, 1975) contrasted with $5 \times 10^{-4}$ in the solar wind. Since the $^{3}$He was completely unaccompanied by $^{2}$H or the elements Li, Be, and B, the $^{3}$He had nothing to do with nuclear fragmentation, rather with a resonance process (*e.g.* Temerin and Roth, 1992) that was associated with streaming electrons and radio type III bursts (Reames, von Rosenvinge, and Lin, 1985; Reames and Stone, 1986). Element enhancements up to Fe (*e.g.* Reames, Meyer, and von Rosenvinge, 1994; Mason, 2007) and subsequently to heavier elements (Reames, 2000; Mason *et al.*, 2004; Reames and Ng, 2004) were apparently accelerated directly in islands of magnetic reconnection (*e.g.* Drake *et al.*, 2009).

However, since $^{3}$He/$^{4}$He varies widely with energy in single events or from event to event (Mason, 2007), while Fe/O exhibits a bimodal distribution with impulsive events enhanced (Reames, 1988), Reames, Cliver, and Kahler (2014a) chose to define impulsive SEP events based upon Fe/O. The distribution in Ne/O *versus* Fe/O near 2.5 MeV amu$^{-1}$ is shown in Figure 1. We define impulsive SEP events as having an abundance of at least four times the reference abundance of Fe/O in the 2.5–3.2 MeV amu$^{-1}$ energy interval.

**Figure 1** Histogram distinguishing impulsive and gradual SEP events shows the occurrence rate of various values of abundances Ne/O *versus* Fe/O in 20 years of 8-hr average measurements from the *Wind* spacecraft. Values which form the centers for selection of impulsive and gradual SEP events by Reames, Cliver, and Kahler (2014a) and Reames (2016a) are indicated. Events defined by Fe/O in this way distinguish meaningful physical processes such as those of associated CMEs.

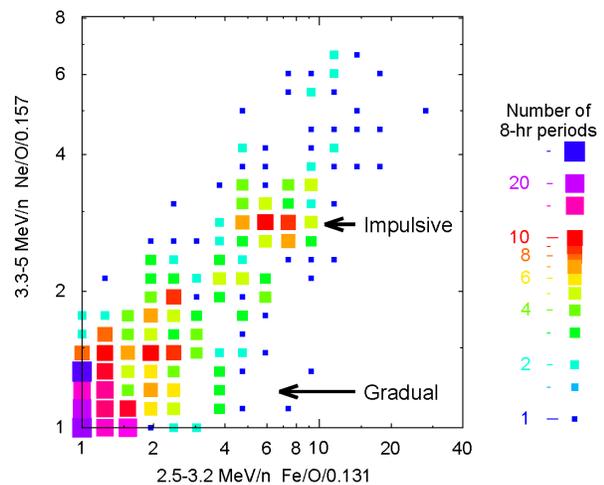

Figure 2 shows properties of two impulsive SEP events that help to illustrate the concepts of "proton excess" and "He suppression." Throughout this article, He without a superscript always means $^{4}$He. Protons in Event 79 barely rise above the pre-event background, yet they do not exceed the power-law fit from the ions with $Z \geq 6$. Of the 111 impulsive SEP events studied (Reames, Cliver, and Kahler, 2014a), 70 have proton intensities above background and 17 (24%) of these have proton intensities within one standard deviation of the value predicted by the best-fit line of the ions with $Z \geq 6$ (Reames, 2019a). Of these 17 events, 6 have order-of-magnitude He suppression. None of the larger events with proton excesses have significant He suppression.

**Figure 2** Intensities (*lower panels*), abundance ratios (*middle panels*) and enhancements *versus A/Q* (*upper panels*) are compared for two impulsive SEP events. Event numbers shown refer to the list of Reames, Cliver, and Kahler (2014a). Event 37 has a proton excess. Event 79 has no proton excess but has a large suppression of He.

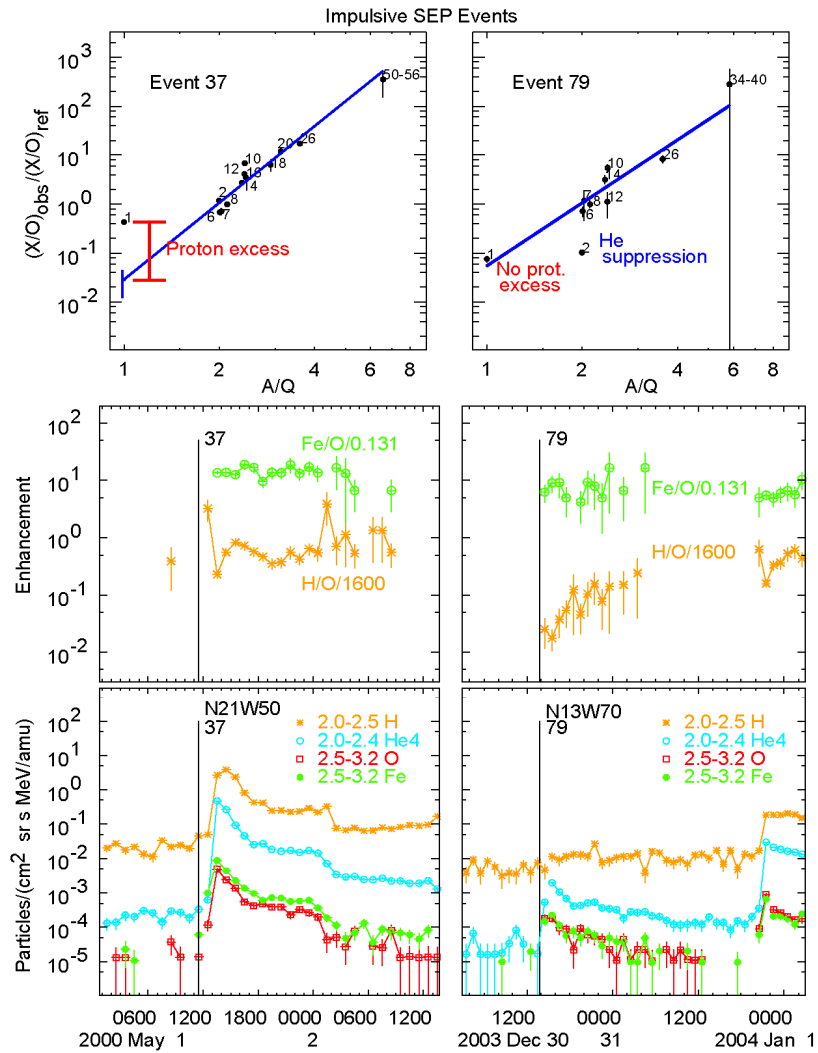

At a temperature of 3 MK, He and C are fully ionized with $Q = Z$ and O is nearly so. Relative abundances of these ions with $A/Q \approx 2$ should remain unaltered during acceleration and transport. Figure 3 shows panels of O/C *versus* He/C with <20% errors, with various quantities represented in the color and size of the points. The scatter in the points is quite large, especially in the He/C direction. Normalization based upon reference values of He/O = 57 and 91 are shown as dashed lines. Various panels show the distribution of temperature, proton excess, peak proton intensity, and CME speed. It is clear that the proton excess is associated with events with higher He/C, with high proton intensities, and with fast CMEs, as noted by Reames (2019a). CMEs with speeds above 500 km s$^{-1}$ are quite capable of ion acceleration. Smaller events do not have proton excess, but may have suppressed He.

**Figure 3**. *Each panel* shows normalized abundances of O/C *versus* He/C with errors <20% for impulsive SEP events with temperature, proton excess, 2 MeV proton intensity, and associated CME speed highlighted by color and size of the points, as indicated. Dashed lines indicate reference abundances of He/O = 57 and 91. Note that the larger events with faster CMEs have higher average He/O.

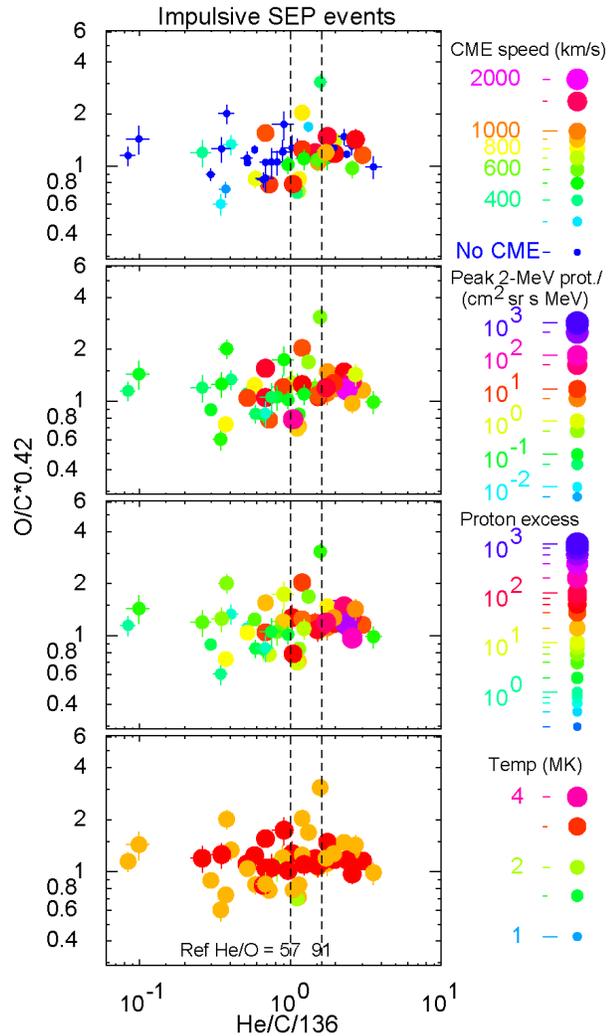

## 3. Gradual SEP Events

Gradual SEP events are best characterized by their strong association with wide, fast CMEs (Kahler *et al.*, 1984; Cane, Reames, and von Rosenvinge, 1988; Reames, Barbier, and Ng, 1996; Gopalswamy *et al.*, 2012; Rouillard *et al.*, 2012; Desai and Giacalone, 2016), an association that is extended by strong CME correlations (Kahler, 2001; Kouloumvakos *et al.*, 2019) and by studies of associated radio type II bursts (*e.g.* Cliver, Kahler, and Reames, 2004).

It is important to emphasize that most gradual SEP events do not show proton excesses. These are events with $T < 2$ MK. A new example is shown in Figure 4 and others are shown by Reames (2019c). Gradual SEP events are analyzed in 8-hr intervals since the abundances often change with time. The best power-law fits are shown for each time interval in the lower-right panel of Figure 4 and the corresponding $\chi^2$ values that select them are shown (color coded) in the upper-right panel. It is not surprising that the fit to protons is occasionally broken, as in the second interval (14 September 2005 at 0000 UT) in Figure 4e, by spatial variations and scattering in the transport path, for example.

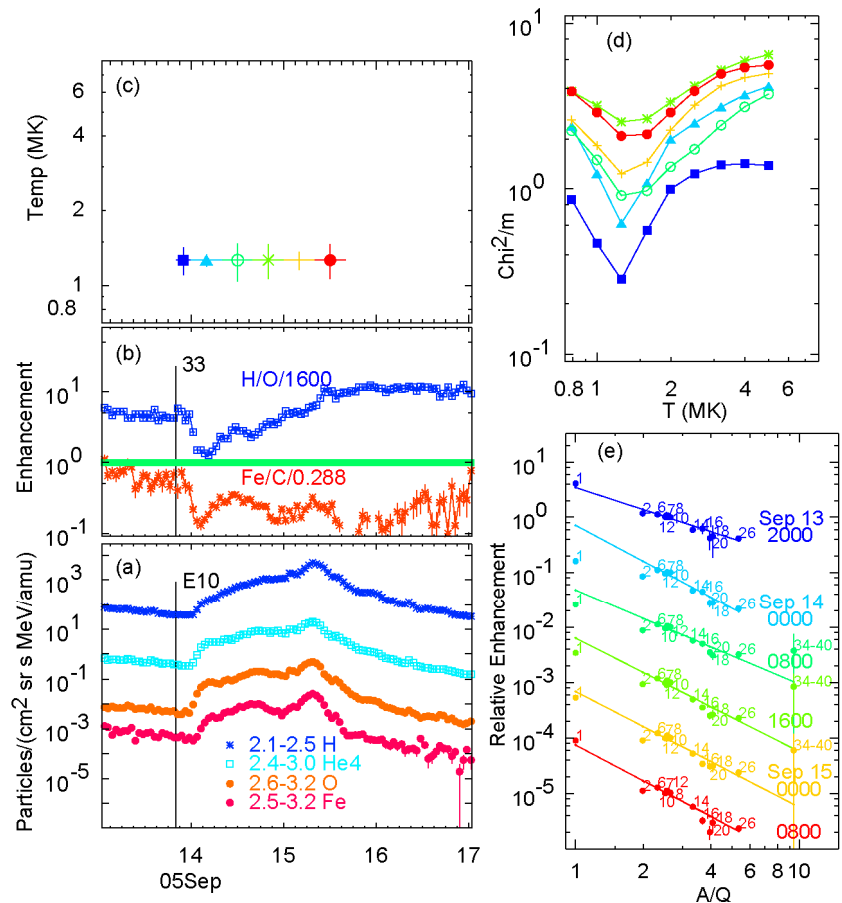

**Figure 4.** (a) Intensities of H, He, O, and Fe *versus* time. (b) Normalized abundance enhancements H/O and Fe/O *versus* time. (c) Best-fit temperatures are shown *versus* time for the 13 September 2005 SEP event. (d) Shows $\chi^2/m$ *versus* $T$ for each 8-hr interval. (e) Shows enhancements, labeled by $Z$, *versus* $A/Q$ for each 8-hr interval shifted ×0.1, with best-fit power law for elements with $Z \geq 6$ extrapolated down to H at $A/Q = 1$. Colors correspond for the six intervals in (c), (d), and (e) and symbols in (c) and (d); times are also listed in (e). Event onset is flagged with solar longitude in (a) and event number 33 from Reames (2016a) in (b). In most intervals in (e), protons fit the power law from higher $Z$.

A clear and persistent proton excess is shown for the event in Figure 5. Power-law distributions for additional gradual SEP events have been shown by Reames (2016a).

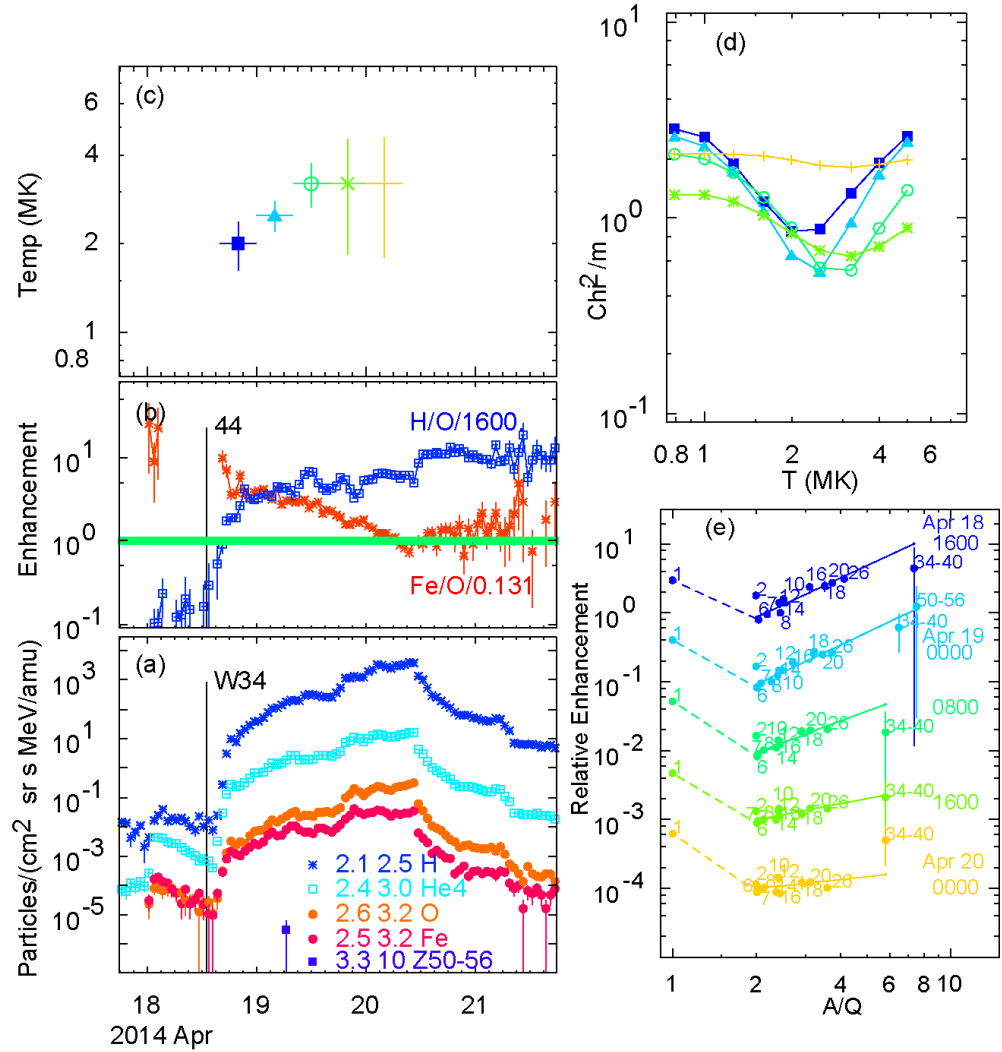

**Figure 5**. (a) Intensities of H, He, O, Fe, and $50 \leq Z \leq 56$ ions *versus* time. (b) Normalized abundance enhancements H/O and Fe/O *versus* time. (c) Temperatures *versus* time for the 18 April 2014 SEP event. (d) Shows $\chi^2/m$ *versus* $T$ for each 8-hr interval. (e) Shows enhancements, labeled by $Z$, *versus* $A/Q$ for each 8-hr interval shifted ×0.1, with best-fit power law for elements with $Z \geq 6$ (*solid*) joined to the enhanced H by *dashed lines*. Colors correspond for the six intervals in (c), (d), and (e) and symbols in (c) and (d); times are also listed in (e). Large proton excesses are seen where dashed lines join H with its associated elements in (e) Event onset is flagged with solar longitude in (a) and event number from Reames (2016a) in (b).

The slope of the power law fitting the enhancements of $Z \geq 6$ ions tends to decrease with time. For the event in Figure 5 this power decreases from 1.9±0.2 to 0.44±0.35 during the five eight-hour periods.  During the final period it becomes so flat that $\chi^2$ also becomes flat and the temperature becomes uncertain.

The lower panel of Figure 6 shows O/C *versus* He/C for gradual SEP events at the same scale as it was shown for impulsive SEP events in Figure 3.  In this case the scale leaves considerable white space because of the tighter grouping of gradual SEP events in O/C *versus* He/C.

**Figure 6** The *lower panel* shows normalized abundances of O/C *versus* He/C for 8-hr intervals during gradual SEP events with temperature as the size and color of the points.  The *upper panels* show Fe/O *versus* He/C with temperature, proton excess, and 20-MeV proton intensity for each interval highlighted by color and size of the points.  Dashed lines indicate reference abundances of He/O= 57 and 91

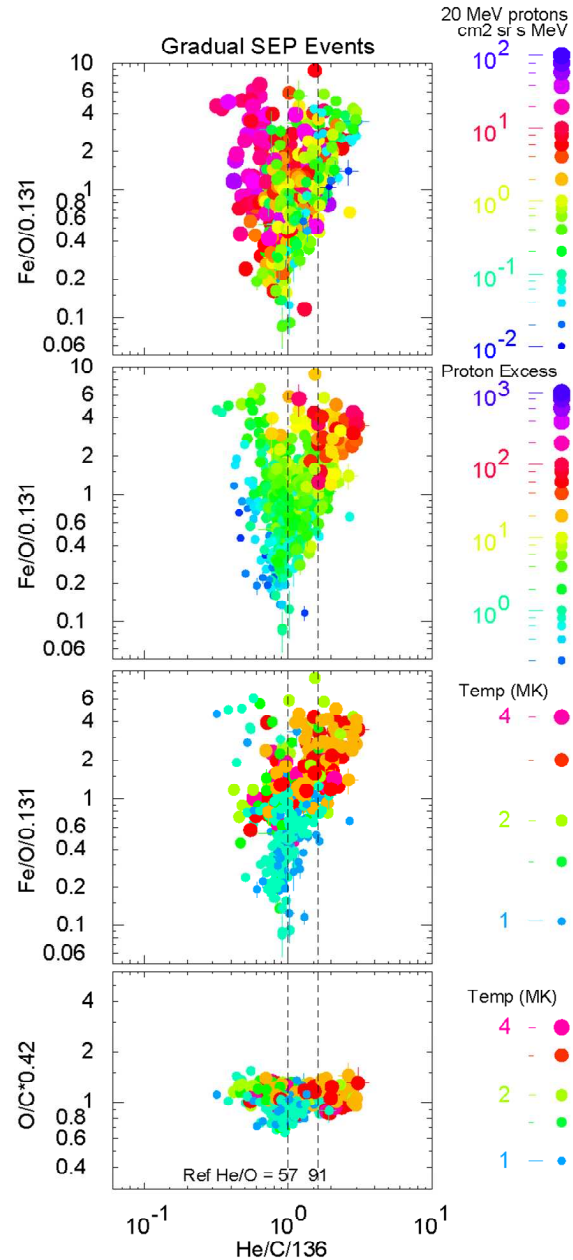

The upper panels in Figure 6 show the tighter grouping of orange and red events with $T \approx 3$ MK in the He-rich, Fe-rich corner of the temperature panels.  Big intense events, as measured by 20-MeV protons and associated with the fastest CMEs (*e.g.* Kahler, 2001; Kouloumvakos *et al.*, 2019), lie along the He-poor side and drift down in Fe/O with time. Thus acceleration in the largest gradual SEP events is dominated by ambient coronal plasma with $T < 2$ MK, not by recycled impulsive material. The gradual SEP events characterized by reaccelerated $T \approx 3$ MK ions from prior impulsive events are modest in size in terms of their energetic proton intensities. Perhaps slower shocks are more strongly biased in favor of pre-accelerated suprathermal ions (Laming *et al.,* 2013).

The shock geometry is also a factor in selecting ions that are accelerated from the seed population (Tylka *et al.*, 2005; Tylka and Lee, 2006). When $\theta_{Bn}$, the angle between the magnetic field and the shock normal, approaches 90°, only the fastest ions can overtake the shock from downstream, so the faster pre-accelerated impulsive ions are favored. Quasi-parallel shocks, with $\theta_{Bn} < 45°$, are less selective.

It was only possible to assign temperatures to about 70% of gradual SEP events since at least four reasonably-consistent eight-hour periods were required (Reames, 2016a). Most events that failed this test did so because the dependence on *A/Q* was too flat, so that the values of *A/Q*, and hence the value of *T*, were indeterminate, *i.e.* there was no "best" value. Since this occurs when the measured abundances are similar to the reference abundances, these events may be less likely to involve many reaccelerated impulsive suprathermal ions with their strong *A/Q* enhancements, and may be biased in favor of sampling ambient coronal material.

## 4. Comparing CMEs

We compare some properties of the CMEs associated with gradual and impulsive SEP events in Figure 7. Relevant CME data were found in the SOHO/LASCO CME catalog (Gopalswamy *et al.*, 2009; https://cdaw.gsfc.nasa.gov/CME_list/). The gradual SEP events on our list have wide, mostly halo, fast CMEs with a mean speed of 1815 km s[-1]. The impulsive event list has narrow, mostly < 120°, slow CMEs with a mean speed for these little CMEs of 687 km s[-1]. For many impulsive events, CME are too small to be recorded (*e.g.* Nitta *et al.*, 2006).

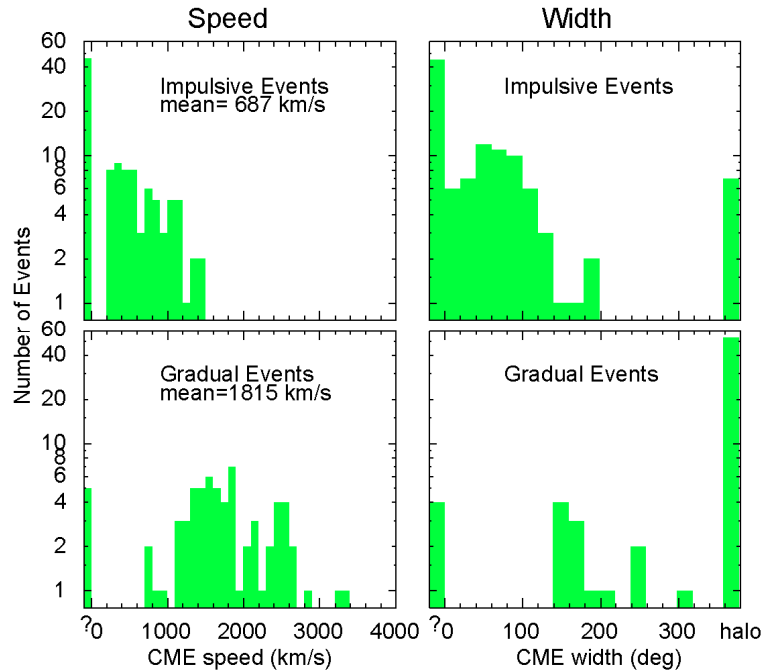

**Figure 7** Histograms compare the speed (*left panels*) and width (*right panels*) distributions of CMEs associated with impulsive (*upper panels*) and gradual (*lower panels*) SEP events. The "?" indicates events with no CME association.

Thus, when we distinguish impulsive and gradual events based upon Fe/O as shown in Figure 1, we find the associated CMEs have differing properties as shown in Figure 7. The properties of the CMEs associated with impulsive SEP events are those of narrow solar jets, those associated with gradual SEPs are large, wide, eruptive CMEs. This leads us to a suggested explanation of the differences of the width of the distribution of He/C, for example, shown in the lower panels of Figures 3 and 6, and first suggested by Reames (2016b) and shown in Figure 8. The abundance patterns on the right in Figure 8 are from Reames (2016b). Both populations involve $T \approx 3$MK impulsive SEP ions from magnetic reconnection sites in solar jets which may be reaccelerated by a shock wave; proton excesses are produced when those shocks also include protons from the ambient coronal material. In the impulsive case, the shock, if any, is produced by the CME from the same event; ions from a single jet reflect local variations in element abundances. In the second case, a large shock from a wide CME sweeps up residual impulsive suprathermal ions from many (N) individual jets that have occurred recently in the vicinity of an entire active region, reducing the abundance fluctuations by a factor of √N. The residue from many jets collects in large regions for substantial periods so that $^3$He-rich, Fe-rich background levels are often seen (Desai *et al.*, 2003; Bučík *et al.*, 2014, 2015, 2018a, 2018b; Chen *et al.*, 2015).

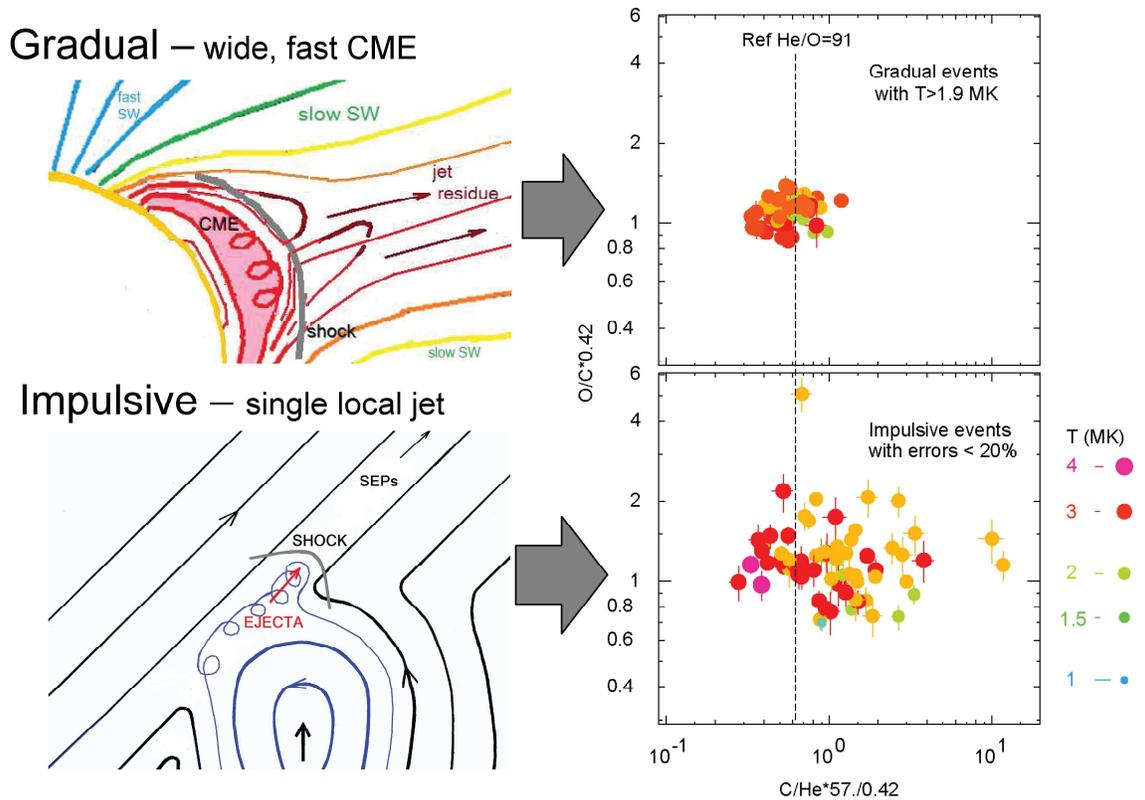

**Figure 8** Suggested explanations for the broad distribution of source abundances in impulsive events (*below*) from variations in single localized jet events where ejecta may or may not drive a shock, and the narrow distribution in high-*T* gradual events (*above*) where pre-accelerated impulsive seed populations from many individual jets are averaged by a large shock. Abundance distributions shown (from Reames, 2016b) are equivalent to those that were shown in the lower panels if Figures 3, and 6.

Having distinguished differences in the CMEs associated with impulsive and gradual SEP events, we now turn to differences in CMEs that favor different ion seed populations for shock acceleration in gradual SEP events. Figure 6 suggests that the most intense gradual events are associated with cooler plasma while events with the hotter reaccelerated impulsive ions are not so large and intense. Figure 9 shows the timing and CME-speed distributions of ions with differing source-plasma temperatures. Events with the hotter reaccelerated impulsive-SEP source plasma involve somewhat slower, weaker CMEs, probably with quasi-perpendicular shock waves, that tend to arrive early in solar cycle 23 and to dominate the weak solar cycle 24. Faster CMEs that occur later in cycle 23 tend to be dominated by cooler ambient coronal plasma in which they sample deeply – these more powerful events have less need for preaccelerated ions. Like SEP events, all solar cycles are not the same.

**Figure 9** shows the CME speed of events as a function of onset time for gradual SEP events, with source plasma temperatures shown, in the *lower panel* and CME speed distributions for gradual SEP events with $T < 1.9$ MK and $T > 1.9$ MK in the *middle* and *upper panels*, respectively.

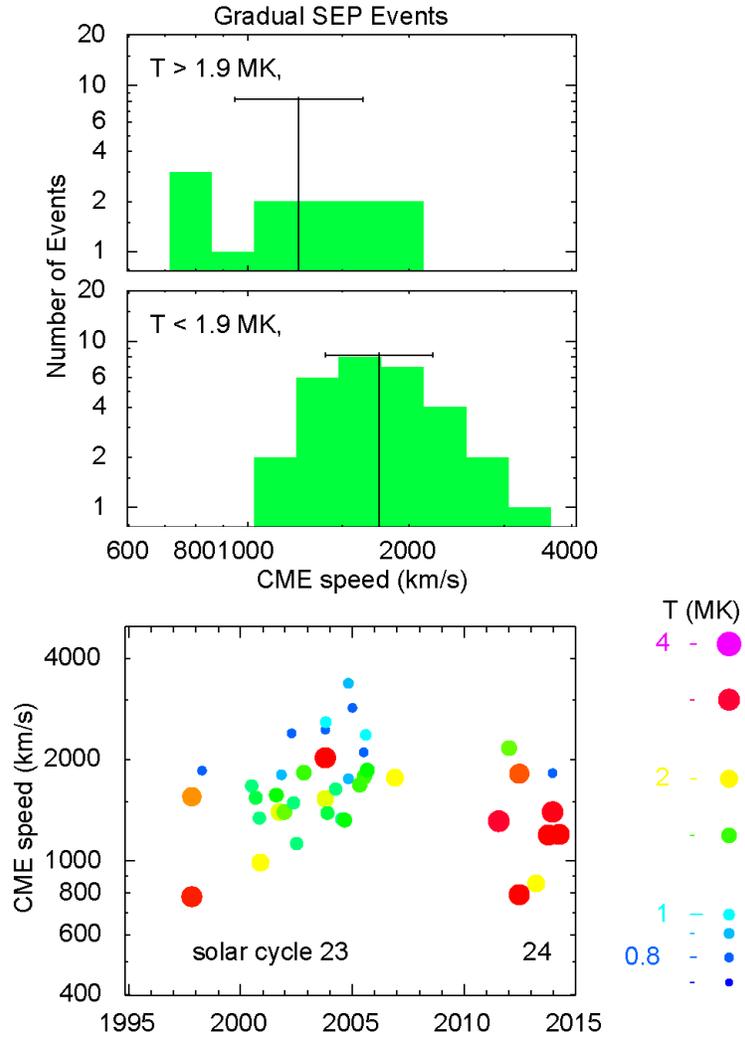

## 5. Discussion

Can the situation described by Figure 8 explain suppressed He? Models of the FIP effect do suggest reduction in the abundance of He because its high value of FIP makes He ionization and fractionation uniquely slow (Laming 2009). However, these calculations show suppression by a factor of two, not a factor of ten. However, these calculations involve a quiescent equilibrium flow of plasma across the chromosphere into the corona. What happens when chromospheric material is suddenly thrust up into the corona in a jet as suggested by the emerging-flux model in the lower-left panel of Figure 8? Would it be surprising for more rapid and dynamic FIP processing to lead to a much greater suppression of He? Perhaps partially FIP-processed chromospheric material may suddenly rise to become involved in some He-poor jets.  However, while $^3$He and $^4$He fractionation should be similar, $^3$He/$^4$He tends to be highest when $^4$He/C is suppressed.

He suppression is the only FIP-related event-to-event variation we can identify in SEP events.  The theory of He suppression (Laming 2009) also suggests the possible suppression of Ne, but Ne is nearly always enhanced in impulsive SEP events (see Figure 1), mainly because of the general increase in enhancement with $A/Q$.  It is possible that the variations in O/C for impulsive events in Figure 8 could occur because O is not quite fully ionized at 2.5 MK so its value of $A/Q$ may be a bit higher than that of C.  Naïve researchers occasionally mistake the enhancement in Fe/O in impulsive SEP events as a possible FIP effect and a direct measure of the FIP bias, but it is actually part of the 1000-fold power law that stretches from H to U (see Figure 8 in Reames, Cliver, and Kahler, 2014a; Reames, 2018b).  This power law in impulsive SEP events also enhances Ne/O, for example, where both elements have high FIP.  The FIP contribution to the coronal abundances seems to be invariant (as contained in the reference in Table 1), it applies as a coronal basis for all SEP events, and shows no evidence of event-to-event FIP variation other than possibly He.  It is important to realize that any FIP-dependent fractionation in impulsive SEP events, including suppression of He, is likely to affect both the SEPs and the ejected plasma from a solar jet; this CME should eventually contribute to the solar wind.  Thus FIP-dependent processes could be sought as SEP-SW correlations, although no such correlations have been reported.

The highest source temperatures, $T \approx 3$ MK for both impulsive and gradual events, offer the most reliable measure of He/O $\approx 90$.  This plasma originally comes from the largest jets which sample material at $\approx 1.5$ solar radii (DiFabio *et al.*, 2008).  Large gradual events sample cooler ambient plasma at $\geq 2$ solar radii (Reames, 2009a, 2009b), but it is the recycled impulsive material in gradual events for which He/O $\approx 90$.  The overall average value for gradual SEP events is He/O $\approx 57$.

For the smaller "pure" (magnetic-reconnection only) impulsive SEP events with no shock reacceleration, and for the largest gradual SEP events which accelerate ambient coronal plasma with $T < 2$ MK, H/O enhancements are generally consistent with power-law fits to the $A/Q$ dependence of abundances of elements with $Z \geq 6$.  Any events domi-nated by shock reacceleration of impulsive suprathermal ions show substantial proton excesses. The only suggested explanation of these proton excesses is shown in Figure 10 where in Event 54 (Reames, Cliver, and Kahler, 2014a; Reames, 2019b) the associated 952 km s[-1] shock wave primarily selects the enhanced impulsive suprathermal ions which dominate the SEPs with $Z > 2$, but protons from the ambient coronal plasma dominate the SEP protons. Abundances of ions with $Z > 2$ rise sharply with $A/Q$ because the impulsive suprathermal source ion abundances already increase sharply with $A/Q$, but the ambient ion abundances and, in fact, all accelerated ion abundances, usually decrease modestly with $A/Q$ because of rigidity-dependent transport from the shock. Note that the cooler ambient ions will have lower $Q$ and higher $A/Q$ than the impulsive ions with $T \approx 3$ MK.

**Figure 10.** Element enhancements, labeled by $Z$, *versus* $A/Q$ for sample impulsive Event 54, with two possible seed-particle sources for shock acceleration from pre-accelerated impulsive ions (*blue*) and from ambient coronal ions or pre-event plasma (*red*). Helium is assumed to receive comparable contributions from both sources for this event (Reames, 2019a). Any weak or quasi-perpendicular shock waves may prefer impulsive suprathermal ions at high Z from a two-component seed-particle source.

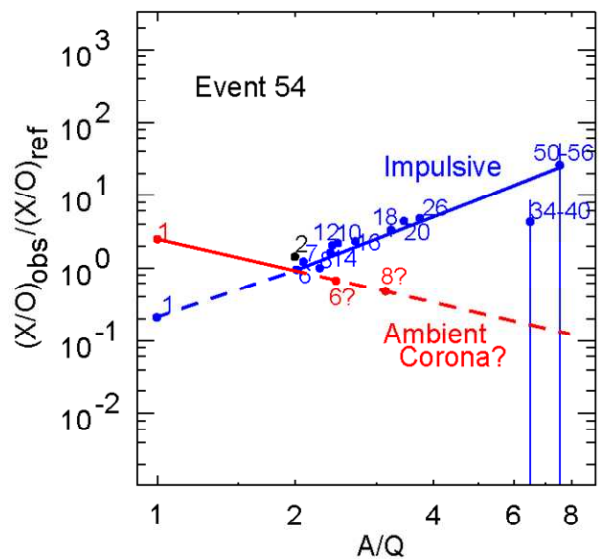

While Figure 10 shows data from an impulsive SEP event, the concept can apply to any event, impulsive or gradual, where a weak or quasi-perpendicular shock wave preferentially reaccelerates residual impulsive suprathermal ions which dominate high $Z$, while protons are mainly sampled from the ambient coronal plasma. However, any ⁴He accelerated from the ambient plasma in impulsive SEP events will certainly tend to reduce ³He/⁴He, so impulsive events with proton excesses may not remain ³He-rich.

Regarding the general issue of multiple components and the seed population, shock acceleration at corotating interaction regions (CIRs) sometimes shares this complexity (Reames, 2018c). CIRs are formed when high-speed solar-wind streams overtake and collide with slow solar wind emitted earlier in the solar rotation (*e.g.* Richardson, 2004).  This collision leads to two shock waves, a forward shock propagating outward into the slow wind and a stronger reverse shock propagating sunward into the fast wind. Usually these shocks form outside 1 AU and strengthen out to 5 AU (Van Hollebeke, McDonald, and von Rosenvinge, 1976).  Low-rigidity ions tend to be retained near the shock while high-rigidity ions spread widely (*e.g.* Reames *et al.*, 1997).  To first order, the only source of particles available for acceleration at CIR shocks is the solar wind, and the energetic particles at CIRs typically show solar-wind abundances near the shocks and abundances that decrease as power-laws in *A/Q* more and more steeply with distance from the shock (Reames, 2018c).  However, seed particles, or simply background, from SEP events, especially small impulsive SEP events, can inject Fe-rich material that easily dominates the otherwise-suppressed abundance of Fe from the solar wind, but contributes much less at lower *Z* (Reames, 2018c).  This is reminiscent of the pattern shown in Figure 10.  These multi-component events tended to hide the *A/Q* dependence and complicated the source composition in early measurements of CIR events, especially for the weaker shocks associated with slower solar-wind streams.

In SEP events, the rigidity dependence that leads to the *A/Q* dependence can come from several factors.  The acceleration of ions in islands of magnetic reconnection leads directly to power laws in *A/Q* when the ions undergo Fermi acceleration as they are re-flected back and forth from the ends of collapsing magnetic islands (Drake *et al.*, 2009). Scattering during transit along magnetic fields also depends upon *A/Q* as noted above (Parker, 1963; Ng, Reames, and Tylka, 2003; Reames, 2016a, 2016b, 2019b).   In shocks, the dominant protons play a special role, not only are they spectators, resonantly scatter-ing from existing waves like all other ions, but they can generate or amplify resonant waves as they stream away from the shock, increasing the scattering and acceleration of ions that follow (Bell, 1978; Lee, 1983, 2005; Ng and Reames, 2008; Reames and Ng, 2010).  The wave number of resonant Alfvén waves is $k \approx B/\mu P$, where *B* is the magnetic field strength, $\mu$ is the cosine of the ion's pitch angle with the field, and *P* is the ion's ri-gidity.  Thus, if we assume $\mu \approx 1$ for simplicity, 2.5 MeV protons are scattered by self-generated waves at 2.5 MeV, for example, but 2.5 MeV amu$^{-1}$ He, C, or O with *A/Q* = 2 (at $T \approx 3$ MK) resonate with waves generated by 10-MeV protons, and 2.5 MeV amu$^{-1}$ Fe at *A/Q* = 4 resonates with waves generated by streaming 39-MeV protons.  A pure power-law dependence of enhancements upon $A/Q$ may require a power-law proton rigidity spectrum. Differences in resonance can produce a proton excess early in a gradual event when, for example, 2.5 MeV protons begin to arrive, as yet unscattered, while 2.5 MeV He at the same time has been scattered and retarded by traversing waves generated by 10-MeV protons that arrived much earlier, increasing H/He (Reames, Ng, and Tylka, 2000). The understanding of SEP events could be greatly enhanced by improved theory and modeling of time dependent generation of resonant waves by protons and the coupling of those waves to other ions. While the full time-dependent evolution of proton and wave spectra at shock waves have been modeled in some detail (*e.g.* Ng and Reames, 2010), the co-evolution of the spectra of other ions has not yet been included.

## 6. Summary

The study of the $A/Q$ dependence of element abundance enhancements has added a new approach to the study of SEP events. Approximate power-law dependence has led to determination of best-fit values of $Q$ and estimates of the underlying source-plasma temperatures. These temperatures distinguish ions enhanced during magnetic reconnection on open field lines in solar jets ($T \approx 3$ MK) from ions of the cooler ($< 2$ MK) ambient coronal plasma. Adding H and He to these abundance studies provides important new information and suggests four categories of acceleration of SEPs:

i) "Pure," shock-free impulsive SEP events accelerate ions in islands of magnetic reconnection in solar jets. Element abundance enhancements increase as a power law in $A/Q$ from H to heavy elements like Pb; they can be distinguished by Fe/O that is over four times the coronal abundance. He/O may be greatly suppressed in occasional events by a rapid rise of material that is too fast to allow much ionization of high-FIP He. Electrons streaming out from the event generate waves that are resonantly absorbed to enhance [3]He; these electrons produce a type III radio burst. Plasma may also be ejected from the event, producing a narrow CME that is too slow to drive a shock wave.

ii) An impulsive + shock event occurs when the narrow CME from a jet, that of an otherwise pure impulsive event, is fast enough to drive a shock wave. The shock wave samples all available ions, those from the ambient plasma and residual energetic ions from the pure event. The abundant ambient protons form the wave structure at the shock and dominate at $Z = 1$, but the pre-enhanced, pre-accelerated ions are selected by the weak shock and dominate the $Z > 2$ region. This appears as a large proton excess.

iii) A fast, wide CME from an eruptive event drives a fast shock wave that expands broadly, producing an energetic gradual SEP event that lasts many days. If the shock is quasi-parallel or samples deeply into the tail of the thermal distribution of the ambient plasma with $T < 2$ MK, it will produce a "pure" gradual event with essentially coronal ion abundances modified by a power-law dependence on $A/Q$ that may be enhanced or suppressed during ion transport. Protons generally fit with other ions although some regions of unusual transport may produce modest local excesses or depletions of protons. Any impulsive suprathermal ions present are also accelerated by the shock, but their contribution is overwhelmed but the accelerated ambient coronal ions; these events do not need pre-accelerated ions.

iv) A fast, wide CME from an eruptive event drives a moderately fast shock wave producing a gradual SEP event. The shock may be quasi-perpendicular (or just weak) so its final contribution from sampling of ambient plasma is limited mainly to protons, while preferring residual impulsive suprathermal ions surviving from a dozen or so earlier impulsive events that combine to produce well-defined average impulsive-SEP abundances for $Z > 2$.

Thus, we still think impulsive SEP events are produced in solar jets, initially from magnetic reconnection, and ions in gradual SEP events are still swept up and accelerated by shock waves driven by wide, fast CMEs. However, weak or quasi-perpendicular shock waves in either event class, manage to accelerate only significant protons from the ambient coronal plasma, but preferentially recycle pre-accelerated impulsive-SEP ions, with their built-in abundance bias, at higher $Z$.

The study of power-law patterns of element abundance enhancements has given us new parameters to better distinguish and organize physical processes in the large sample of SEP events we have seen. Soon the *Wind* spacecraft will enter its third solar cycle;

will the abundances of the new SEP events return to the pattern of solar cycle 23, as we would predict, or will they differ entirely?

**Acknowledgements** CME data were taken from the CDAW LASCO catalog. This CME catalog is generated and maintained at the CDAW Data Center by NASA and The Catholic University of America in cooperation with the Naval Research Laboratory.

# Disclosure of Potential Conflicts of Interest

The author declares he has no conflicts of interest.

# Appendix: Reference Abundances of Elements

The average element abundances in gradual SEP events are a measure of the coronal abundances sampled by SEP events (Reference SEPs in Table 1). They differ from photospheric abundances in Table 1 by a factor which depends upon FIP (*e.g.* Reames, 2018a; 2018b; Laming *et al.*, 2019). More complete tables of SEP abundances of 21 elements that include rarer species are given by Reames (2017a; 2018a). Ion "enhancements" are defined as the observed abundance of a species, relative to O, divided by the reference abundance of that species, relative to O.

**Table 1** Reference SEP abundances, photospheric, and impulsive SEP abundances are compared for various elements.

| | Z | FIP [eV] | Photosphere[1] | Reference SEPs[2] | Avg. Impulsive SEPs[3] |
|---|---|---|---|---|---|
| H | 1 | 13.6 | $(1.74\pm0.17)\times10^{6}$ [*] | $(1.6\pm0.2)\times10^{6}$ | - |
| He | 2 | 24.6 | 146000±6700 | 57000±5000 | 53000±3000 [**] |
| C | 6 | 11.3 | 550±76 [*] | 420±10 | 386±8 |
| N | 7 | 14.5 | 126±35 [*] | 128±8 | 139±4 |
| O | 8 | 13.6 | 1000±161 [*] | 1000±10 | 1000±10 |
| Ne | 10 | 21.6 | 195±45 | 157±10 | 478±24 |
| Mg | 12 | 7.6 | 60.3±8.3 | 178±4 | 404±30 |
| Si | 14 | 8.2 | 57.5±8 | 151±4 | 325±12 |
| S | 16 | 10.4 | 25.1±2.9 [*] | 25±2 | 84±4 |
| Ar | 18 | 15.8 | 5.5±1.3 | 4.3±0.4 | 34±2 |
| Ca | 20 | 6.1 | 3.7±0.6 | 11±1 | 85±4 |
| Fe | 26 | 7.9 | 57.5±8.0 [*] | 131±6 | 1170±48 |
| Se-Zr | 34-40 | - | ≈0.0118 | 0.04±0.01 | 2.0±0.2 |
| Sn-Ba | 50-56 | - | ≈0.00121 | 0.0066±0.001 | 2.0±2 |
| Os-Pb | 76-82 | - | ≈0.00045 | 0.0007±0.0003 | 0.64±0.12 |

[1] Lodders, Palme, and Gail (2009), see also Asplund *et al.* (2009).

[*] Caffau *et al.* (2011).

[2] Reames (1995a, 2014, 2017a).

[3] Reames, Cliver, and Kahler (2014a)

[**] Reames (2019a).